# Energy Efficient Resource Allocation in Vehicular Cloud based Architecture

Amal A. Alahmadi, Mohammed O. I. Musa, T. E. H. El-Gorashi, and Jaafar M. H. Elmirghani
*School of Electronic & Electrical Engineering, University of Leeds, LS2 9JT, United Kingdom*
*e-mail: {elaaal, m.musa, t.e.h.Elgorashi, j.m.h.elmirghani}@leeds.ac.uk*

**ABSTRACT**
The increasing availability of on-board processing units in vehicles has led to a new promising mobile edge computing (MEC) concept which integrates desirable features of clouds and VANETs under the concept of vehicular clouds (VC). In this paper we propose an architecture that integrates VC with metro fog nodes and the central cloud to ensure service continuity. We tackle the problem of energy efficient resource allocation in this architecture by developing a Mixed Integer Linear Programming (MILP) model to minimize power consumption by optimizing the assignment of different tasks to the available resources in this architecture. We study service provisioning considering different assignment strategies under varying application demands and analyze the impact of these strategies on the utilization of the VC resources and therefore, the overall power consumption. The results show that traffic demands have a higher impact on the power consumption, compared to the impact of the processing demands. Integrating metro fog nodes and vehicle edge nodes in the cloud-based architecture can save power, with an average power saving up to 54%. The power savings can increase by 12% by distributing the task assignment among multiple vehicles in the VC level, compared to assigning the whole task to a single processing node.
**Keywords**: mobile edge computing (MEC); vehicular clouds; fog; resource assignment; resource allocation; service provisioning; power consumption.

## 1. INTRODUCTION

Providing energy-efficient and reliable network infrastructure has been the focus of attention in recent years. The huge growth in the Internet traffic and the growth of cloud-dependant applications have had a large impact on the network power consumption. This calls for more research to develop new architectures and solutions to provide robust and energy-efficient infrastructures. Previous research efforts contributed solutions to reduce the power consumption of cloud datacenters and core networks [1] - [5], considering many research approaches including virtualization [6], [7], network architecture design and optimization [8]-[12], content distribution [13], big data [14] - [16], network coding [17], [18] and renewable energy [19]. Decentralized architectures have also been proposed to integrate distributed edge servers to mitigate the traffic burden on central data centers [20], [21].

Vehicular clouds (VC) are one of the distributed computing paradigms that has been proposed recently. They exploit the powerful and underutilized on-board computation and storage resources in vehicles in a cloud computing fashion. Integrating VC with centralized cloud services can enhance the service provisioning by serving users locally. This integration can also offload the computational burden and reduce the power consumption in conventional data centers. Recent research efforts have been successful in adopting the vehicles in cloud computing paradigm as static vehicles in a parking lot [22], [23], moving vehicles [24], [25], or vehicles stopping at traffic lights [26]. However, the hybrid architecture and dynamic behavior of the VC make it a very interesting direction with many challenging areas, including heterogeneous resource management [27], service provisioning [28], quality of experience [29], and resource scheduling and allocation [30], [31]. The problem of resource allocation can be a major factor affecting the total network power consumption, especially in the presence of edge computing. Optimization approaches have been studied in vehicular cloud environments to achieve the optimal scheduling strategy in term of execution delay [32], continuous service [33], resource utilization [34], [35], and cost minimization [36] – [38].

In this paper, we extend our previous work in [26], (where the main focus was minimizing the conventional cloud power consumption), by introducing a vehicular cloud to the architecture. We tackle the problem of energy cost minimization by optimizing resource assignment in an architecture that is a composite of a cluster of vehicles' on-board units referred to as a vehicular cloud, integrated with metro fog servers and a conventional cloud datacenter. We also study and compare different resource assignment strategies in terms of delivering the best overall network power savings. The remainder of this paper is organized as follows: In Section 2, the proposed VC architecture is presented and the energy-efficient resource allocation strategies are explained. The optimization results are presented and discussed in Section 3. Finally, Section 4 concludes the paper.

## 2. VEHICULAR CLOUD ARCHITECTURE AND ENERGY EFFIEICNT ASSIGNMENT STRATEGIES

Figure 1 shows the proposed architecture which is comprised of the conventional cloud servers, the metro fog nodes, and vehicles assumed to be clustered at an intersection near a traffic light. Vehicles at each intersection communicate with an access point attached to the traffic light. Tasks originating from nearby devices are collected by the roadside unit (RSU) which has sufficient computational capability to execute the assignment strategy. Based on the executed strategy, the RSU offloads each task either to the conventional cloud, fog nodes, or connected vehicles.

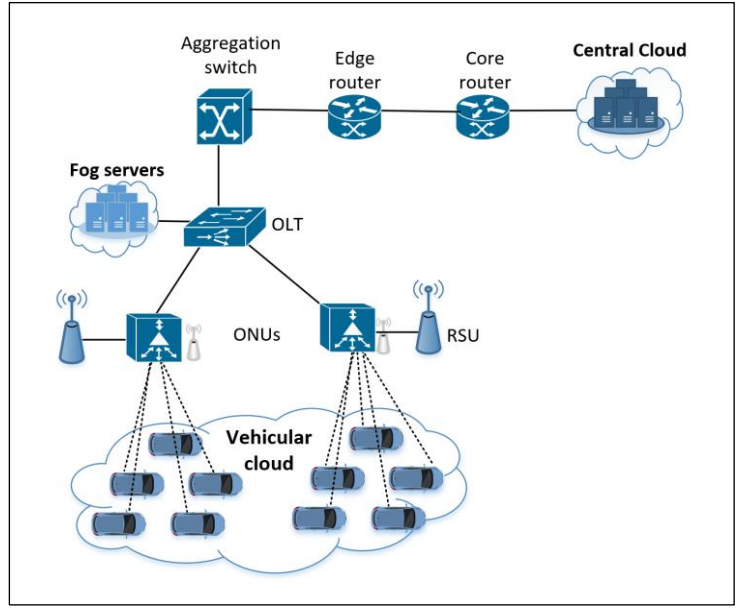

*Figure 1. Integrated Vehicular Cloud Architecture*

The tasks assignment is optimized using mixed integer linear programming to minimize the power consumption of the overall architecture. The model objective is to minimize the power consumption of the overall proposed architecture including the processing and networking power consumption of the cloud servers, fog nodes, VC OBUs, and the intermediate optical and wireless links.

Five different assignment strategies are considered. All strategies are evaluated under a varying traffic demand volume as well as a varying processing demand volume. These assignment strategies are defined as follows:

- ***Conventional Cloud-based assignment (Cloud):*** In this strategy, all tasks are assigned only to the conventional cloud. This represents the baseline approach to which assignment strategies considering the VC are compared. This is executed in the model by introducing a constraint that restrict the assignment only to the conventional cloud.
- ***Optimal Cloud/Fog based assignment (C/F-optimal):*** In this strategy, tasks are assigned to the conventional cloud, fog servers, or split between the cloud and fog. This assignment is optimized by the MILP model to guarantee the minimum power consumption of the architecture. This represents another baseline approach to which assignment strategies (considering the VC) are compared.
- ***Optimal Cloud/Fog/VC based single assignment (C/F/V-single):*** In this strategy, each task is assigned to one location: the Cloud, the Fog, or the VC based on the minimum overall power consumption. This assignment, however, uses single assignment where no splitting among servers takes place. This is enforced in the MILP by using a no splitting constraint among candidate processing nodes.
- ***Optimal Cloud/Fog/VC based distributed assignment (C/F/V-distributed):*** In this strategy, each task is assigned to one or more location in the Cloud, Fog, or VC based on the minimum power consumption objective. The model is free to split the tasks to achieve this objective. When distributing the processing tasks, the related traffic is split among processing destinations proportionally.
- ***Cloud/Fog/VC based random assignment (C/F/V-random):*** In this strategy, tasks are assigned to only one location randomly between Cloud, Fog, and VC. The RSU offloads the task to the randomly chosen location of the three processing locations. However, if tasks assigned to the VC exceed its capacity, these tasks will be offloaded to the fog node. Random assignment has been widely adopted to analyze network performance, and it can work as a baseline for other assignment strategies of the Cloud/Fog/VC architecture.

## 3. RESULTS AND DISCUSSION

We evaluate the different assignment strategy in a scenario of 20 vehicles with homogeneous OBUs, 15 fog servers and 50 requested tasks. Based on the considered servers, the processing energy of vehicle's OBU is 90% more efficient compared to the cloud servers processing energy. Followed by fog nodes, with 52% more energy efficiency compared to the cloud. Table 1, summarizes the processing capacity and power network of components in the different layers in the architecture.

In this section, we study the impact of increasing the traffic and processing demands on the energy efficiency of the overall architecture, considering the assignment strategies described in the previous section.

*Table 1. Network components parameters*

| Network Components | Capacity | Power consumption | Network Components | Capacity | Power consumption |
|---|---|---|---|---|---|
| Conventional server | 4 GHz | 300 W [39] | OLT (Tellabs1134) | 320 Gb/s | 400 W [40] |
| Cloud switch (cisco 6509) | 320 Gb/s | 3.8 kW [39] | ONU (Tellabs ONT140C) | 2.488 Gb/s | 5 W [41] |
| Cloud router (Juniper MX-960) | 660 Gb/s | 5.1 kW [39] | Access point (Extreme 3825i/e) | 1.75 Gb/s | 7.42 W [42] |
| Core router (Cisco CRS 16-slots) | 12.8 Tb/s | 13.9 kW [43] | RSU (Savari SW-1000) | 27 Mb/s | 7 W [44] |
| Optical switch (Cisco SG220-50P) | 100 Gb/s | 63.2 kW [45] | Low-end computer (Intel Atom) | 1.6 GHz | 18 W [46] |
| Transponder (Cisco ONS 15454) | 10 Gb/s | 50 W [47] | OBU processor (AMOS-825) | 1 GHz | 7 W [48] |
| Edge router (Cisco 12816) | 200 Gb/s | 4.2 kW [49] | Vehicle network Wi-Fi (CC3000 Module) | 54 Mb/s | 207mW [50] |
| Aggregation switch (Cisco 6880) | 160 Gb/s | 3.8 kW [51] | Fog server (Intel Core2-Q9400) | 2.66 GHz | 95 W [52] |

*A. The impact of traffic demands (data size) on the assignment decision and power consumption*

The processing demand of tasks follows a normal distribution with mean value equal to 1GHz and standard deviation equal to 0.5GHz. The traffic of the task is normally distributed and vary the between mean values of 10-100Mb/s and standard deviation 5Mb/s.

As shown in Fig. 2a, the total power consumption in the cloud based assignment follows a linear relationship with the traffic demands of tasks, and this applies to C/F and C/F/V distributed assignment strategies.

The random and single assignment strategies follow a non-linearly increasing trend in the power consumption. This is due to the limitation of wireless link capacity in the VC which leads to offloading tasks with traffic demands that exceed the link capacity to the fog node and conventional cloud. Fig.3 shows that the random assignment achieves 30% power savings compared to the cloud assignment, but results in 7% power loss compared to the optimal cloud/fog based assignment. This is because, the random assignment chooses randomly between the Cloud, Fog, and the VC. This makes the random assignment less efficient than the optimal cloud/Fog strategy which prioritizes the use of Fog nodes. On the other hand, the single assignment achieves an average power saving of 45% and 18% compared to the cloud based and F/C based assignment strategies, respectively.

Similar to the Cloud and C/F assignment strategies, the power consumed in the distributed assignment follows a linear relationship with the traffic demands of tasks. However, the distributed assignment achieves the highest power savings, with averages of 54%, 30%, and 12% compared to the cloud based, F/C based, and single assignment strategies, respectively. This is due to its ability to split the tasks and utilize the VC resources by assigning more demands to the energy efficient VC's OBUs.

Figure.3 shows the average power savings among the different strategies compared to the cloud based and C/F based strategies.

*B. The impact of processing demands on the assignment decision and power consumption:*

In this evaluation, the traffic demand of a task follows a normal distribution with mean value equal to 50 Mb/s and standard deviation equal to 5Mb/s.. The processing demand is normally distributed and vary the between mean values of 0.1GHz and1GHz and standard deviation 0.5GHz.

Fig.2b shows that the power consumption of the conventional cloud based assignment follows a linear relationship with the processing demand of tasks. Furthermore, in the Cloud/Fog based assignment, the power increases linearly with increase in the processing demands until the fog capacity is exhausted, and a step power increase occurs afterwards with a higher rate.

Integrating the VC to the architecture while considering the random assignment reduces the power consumption by an average of 30% compared to the conventional cloud based assignment. However, random assignment results in 11% power loss compared to the optimal cloud/fog assignment, as shown in Fig.4b. This is because, the random assignment chooses randomly between the Cloud, Fog, and the VC and therefore the less efficient cloud resources get used in many cases.

Under the single and distributed assignment strategies, the power consumption increases linearly with increase in the processing demands until the VC capacity is exhausted, and a power increase occurs afterwards with a higher rate. The single assignment achieves average power savings of 46% and 11%, compared to the cloud based and C/F based strategies, respectively. Comparable to this saving, the distributed assignment reduces the power consumption by an average of 47% and 21% compared to the cloud based and C/F based strategies, respectively. Average power savings comparsions are presented in Fig.4.

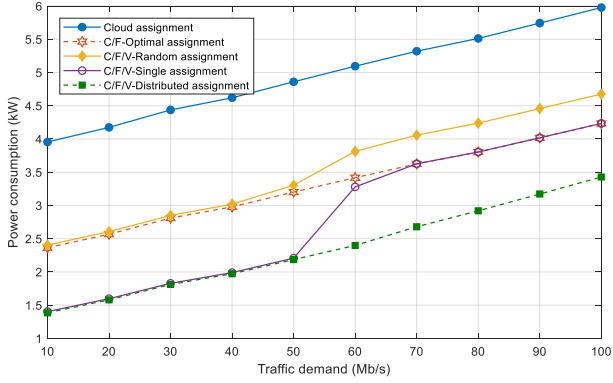
(a) Impact of traffic demand

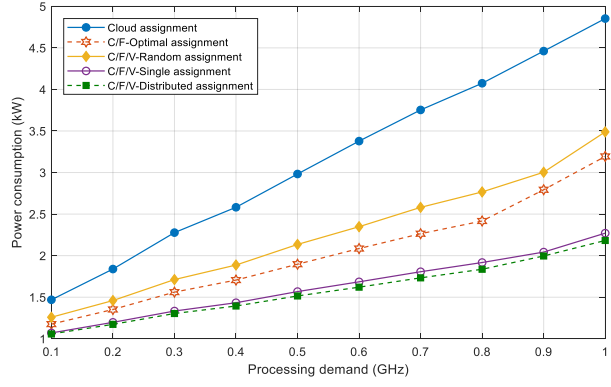
(b) Impact of processing demand

*Figure 2. Impact of increasing the demand on the power consumption for different assignment strategies*

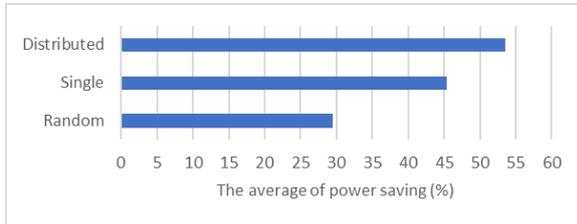
(a) Power savings in C/F/V based assignment strategies, compared to Conventional cloud assignment

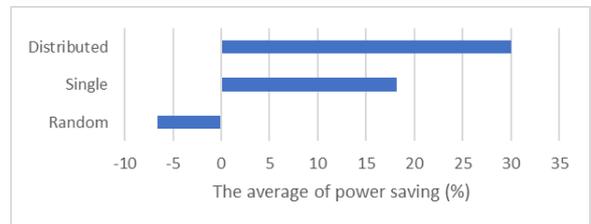
(b) Power savings in C/F/V based assignment strategies, compared to Cloud /Fog-Optimal assignment

*Figure 3. The average power savings when increasing traffic demand*

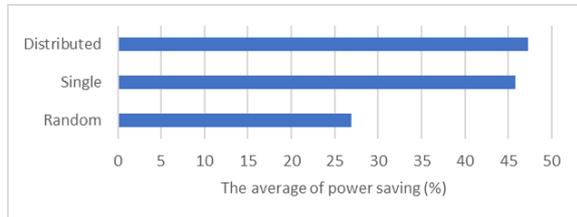
(a) Power savings in C/F/V based assignment strategies, compared to Conventional cloud assignment

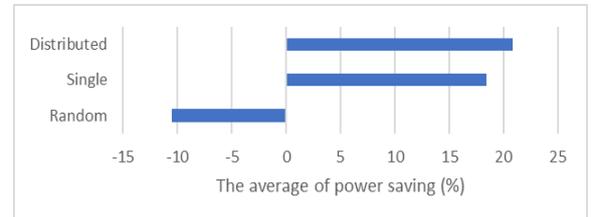
(b) Power savings in C/F/V based assignment strategies, compared to Cloud/Fog-Optimal assignment

*Figure 4. The average power savings when increasing processing demand*

## 4. CONCLUSIONS

This study proposed a distributed MEC architecture which is comprised of the a conventional cloud, metro fog nodes, and vehicles clustered by a traffic light. The proposed architecture was modelled using MILP to solve the problem of energy efficient resource assignment. Five assignment strategies were evaluated while varying the processing and traffic demands of the generated tasks. Our results showed that the traffic volume has higher impact on the power consumption compared to the impact of the processing demands. The results show an average saving between 30% and 54% when fog nodes and vehicles OBUs are integrated to the cloud based architecture. The results show a priority in terms of task allocation firstly in the VC, followed by metro fog nodes, and finally the conventional cloud. This is due to the fact that the energy needed for processing in on board units is 90% lower compared to the energy needed in cloud servers, followed by fog nodes, with 52% more energy efficiency compared to the cloud. Dividing the tasks to more than one server to be processed has a better power saving compared to the single task assignment.

**ACKNOWLEDGEMENTS**

The authors would like to acknowledge funding from the Engineering and Physical Sciences Research Council (EPSRC), through INTERNET (EP/H040536/1), STAR (EP/K016873/1) and TOWS (EP/S016570/1) projects. All data are provided in full in the results section of this paper.